\documentclass[aps,preprint,groupedaddress,superscriptaddress]{revtex4}

\usepackage{graphicx}
\usepackage{dcolumn}
\usepackage{bm}
\usepackage{multirow}
\usepackage{color}
\usepackage{amsmath}
\usepackage{amssymb}

\begin{document}

\author{Jeonghwan Ahn$^{\ast}$}
\affiliation{Materials Science and Technology Division, Oak Ridge National Laboratory, Oak Ridge, TN 37831, USA}
\author{Seoung-Hun Kang$^{\ast}$}
\affiliation{Materials Science and Technology Division, Oak Ridge National Laboratory, Oak Ridge, TN 37831, USA}
\author{Mao-Hua Du}
\affiliation{Materials Science and Technology Division, Oak Ridge National Laboratory, Oak Ridge, TN 37831, USA}
\author{Mina Yoon}
\affiliation{Materials Science and Technology Division, Oak Ridge National Laboratory, Oak Ridge, TN 37831, USA}
\author{Jaron T. Krogel}
\affiliation{Materials Science and Technology Division, Oak Ridge National Laboratory, Oak Ridge, TN 37831, USA}
\author{Fernando A. Reboredo$^{\dag}$}
\affiliation{Materials Science and Technology Division, Oak Ridge National Laboratory, Oak Ridge, TN 37831, USA}
\email{reboredofa@ornl.gov}

\title{Procedures for assessing the stability of proposed topological materials}



\begin{abstract}
We investigate the stability of MnPb$_{2}$Bi$_{2}$Te$_{6}$ (MPBT), which is predicted to be a magnetic topological insulator (TI), using density functional theory calculations. Our analysis includes various measures such as enthalpies of formation, Helmholtz free energies, defect formation energies, and dynamical stability. Our thermodynamic analysis shows that the phonon contribution to the energy gain from finite temperature is estimated to be less than 10~meV/atom, which may not be sufficient to stabilize MPBT at high temperatures, even with the most favorable reactions starting from binaries. While MPBT is generally robust against the formation of various defects, we find that anti-site defect formation of $\text{Mn}_{\text{Pb}}$ is the most likely to occur, with corresponding energy less than 60~meV. This can be attributed to the significant energy cost from compressive strain at the PbTe layer. Our findings suggest that MPBT is on the brink of stability in terms of thermodynamics and defect formation, underscoring the importance of conducting systematic analyses of the stability of proposed TIs, including MPBT, for their practical utilization. This study offers valuable insights into the design and synthesis of desirable magnetic TI materials with robust stabilities.
\end{abstract}

\newpage
\maketitle

$^\ast$ These authors contributed equally to this work.

$^\dag$ reboredofa@ornl.gov

\section{Introduction}

The quantum spin Hall effect~\cite{zhang2009topological,xia2009observation,hsieh2009observation,chen2009experimental} exhibited by topological insulators (TI) has captured the attention of researchers over the years. This phenomenon results in the flow of a spin-polarized current without any resistance at the surface due to a protected Dirac state by spatial and time-reversal symmetry~\cite{hasan2010colloquium,qi2011topological}. Researchers have recently explored exotic quantum phases, such as those exhibiting the quantum anomalous Hall effect (QAHE)~\cite{chang2013experimental,chang2015high,deng2020quantum} and axion insulator states~\cite{mogi2017magnetic,mogi2017tailoring,liu2020robust}, by breaking time-reversal symmetry through magnetism. To accomplish this, magnetically doped TIs~\cite{chang2013experimental,chang2015high} or heterostructures combining TIs and magnetic compounds~\cite{mogi2017magnetic,mogi2017tailoring} have been proposed. However, synthesizing these materials is a challenging process as it requires precise control to obtain interfaces~\cite{lee2015imaging,dai2016toward} with uniformly distributed constituent elements. Alternatively, a family of TIs that includes ternary or quaternary variants, whose parent materials are the well-known binary TI materials, provides another avenue to achieve a single-material platform for studying emergent quantum phases associated with magnetism.

MBT, the first intrinsic magnetic TI, was synthesized by combining Bi${2}$Te${3}$ and MnTe to exploit the latter's intrinsic magnetism~\cite{otrokov2019prediction}. As a platform to realize the Majorana zero mode, A-type magnetic MBT has shown potential due to the interplay between topology and magnetism, which is highly desirable for topological quantum computing~\cite{otrokov2019prediction,li2019intrinsic}. Recently, research on MBT has shifted towards exploring its magnetic topological states and functionalities by manipulating its magnetism and band topology through chemical doping~\cite{chen2019intrinsic,yan2019evolution,Qian2022}, structural engineering by applying pressure~\cite{chen2019suppression,shao2021pressure,qian2022unconventional}, and adding a Bi${2}$Te${3}$ layer~\cite{aliev2019novel,wu2019natural,hu2020realization}. MPBT, a theoretical material that could be achieved through doping Pb atoms onto MBT, has been predicted to also be a magnetic TI~\cite{gao2022intrinsic,gao2022intrinsic2,Qian2022}. By introducing heavy element Pb into MBT, MPBT is expected to show an enhanced band gap, as well as much reduced magnetic exchange coupling between Mn layers due to its expanded unit structure from septuple to undecuple layer. These features make MPBT a promising candidate for practical applications of magnetic TIs. However, the synthesis of MPBT is likely to be challenging due to its complex elemental composition and the potential for defects arising from different strain-field relaxations. This aspect has not been thoroughly addressed for MPBT, and the robustness of the material against different measures of stability remains uncertain. Additionally, MBT is known to be difficult to synthesize as a single crystal due to high-temperature metastability~\cite{zeugner2019chemical,lee2013crystal} and various defects~\cite{zeugner2019chemical,huang2020native,lai2021defect}. Therefore, further research is necessary to fully understand the properties and potential applications of both MBT and MPBT.

In this study, we first evaluated the thermodynamic stability of MPBT for various reactions and defect formations using multiple computational schemes. Our results indicate that the synthesis of MPBT from binary compounds is expected to result in a small energy gain at finite temperatures. Additionally, we estimated the formation energy for various intrinsic point defects and found that $\text{Mn}{\text{Pb}}$ is the most likely defect to form due to the energy cost associated with compressive strain at the PbTe layer. These findings suggest that MPBT is on the edge of stability with regard to both thermodynamics and defect formation. This is reminiscent of the situation with MBT and highlights the need for caution when proposing new TI materials based on ab {\it initio} calculations. While the general picture of stability in MPBT is not significantly affected by the computational scheme, the formation energy of $\text{Mn}{\text{Pb}}$ is spread near the stabilization of MPBT, depending on the specific computational approach used.

\section{Computational Methods and Details}

We employed density functional theory (DFT)\cite{{Hohenberg1964},{Kohn1965}} with the Vienna \textit{ab initio} simulation package (VASP)\cite{{Kresse1993},{Kresse1996}} to perform \textit{ab initio} calculations. The calculations incorporated spin-orbit coupling (SOC) and used projector-augmented wave potentials~\cite{{PAW1994},{Kresse1999}}. We employed the Perdew-Burke-Ernzerhof (PBE) form~\cite{Perdew1996} for the exchange-correlation functional within the generalized gradient approximation and the DFT-D3 correction method~\cite{grimme10} to handle vdW interactions. The energy cutoff was set to 350eV for all calculations with the optimized structure. We used a $3\times3\times1$ supercell with a 13.25\AA{} in-plane lattice constant optimized with the PBE+D3 calculations to reduce interactions between defects in neighboring cells. The structural relaxations had an accuracy of 10$^{-2}$~eV/\AA. A $15\times15\times3$ $\Gamma$-centered k-grid was used to sample the Brillouin zone in a pristine system. We employed the Dudarev scheme~\cite{Dudarev1998} for DFT+$U$ calculations. The value of $U$ was obtained from Diffusion Monte Carlo (DMC) calculations of MBT previously performed with QMCPACK\cite{kim2018,kent2020}, resulting in a value of 3.5(2)~eV, selected using the fixed node variational principle to minimize the DMC total energy. This value has yielded a good description of magnetic properties in correlated materials. We applied Hubbard repulsions of $U$ = 3, 3.5, and 4~eV to the Mn 3$d$ states on the same structure optimized with PBE+D3 calculations to assess the effects of uncertainty in $U$ and SOC on our results, independent of the structural degrees of freedom. Using $U$ = 3 $\sim$ 4~eV is also consistent with the range of values commonly used in previous studies~\cite{du2021tuning,bennett2022magnetic,lupke2022local,gao2022intrinsic,gao2022intrinsic2}.

To evaluate the dynamical stability of MPBT and its Helmholtz free energy as a function of temperature, we used density functional perturbation theory (DFPT)\cite{baroni2001phonons} within the quasi-harmonic approximation (QHA). Specifically, we calculated the phonon spectra using the QUANTUM ESPRESSO package\cite{giannozzi09}. The Helmholtz free energy is given by 
\begin{equation}
 F(V,T) = U(V) + \frac{1}{2}\sum_{{\bf q},j}\hbar\omega_{j}({\bf q}) + k_{B}T\sum_{{\bf q},j}ln[1-e^{-\frac{\hbar\omega_{j}({\bf q})}{k_{B}T}}]
\label{eqn:areaset}
\end{equation} 
where $U$ represents the internal energy from the DFT total energy, while the second and third terms correspond to the zero-point vibrational and thermal contributions, respectively. The phonon frequencies, denoted by $\omega_{j}({\bf q})$, were computed for each mode at the wave vector ${\bf q}$. We constructed the dynamical matrices using force constant matrices computed on a $4\times4\times3$ ${\bf q}$-grid. The optimized norm-conserving Vanderbilt pseudopotential\cite{hamann2013} was used for the phonon calculations, with a plane wave energy cutoff of 80~Ry.

\section{Results}

\begin{figure}
\centering
\includegraphics[width=2.8in]{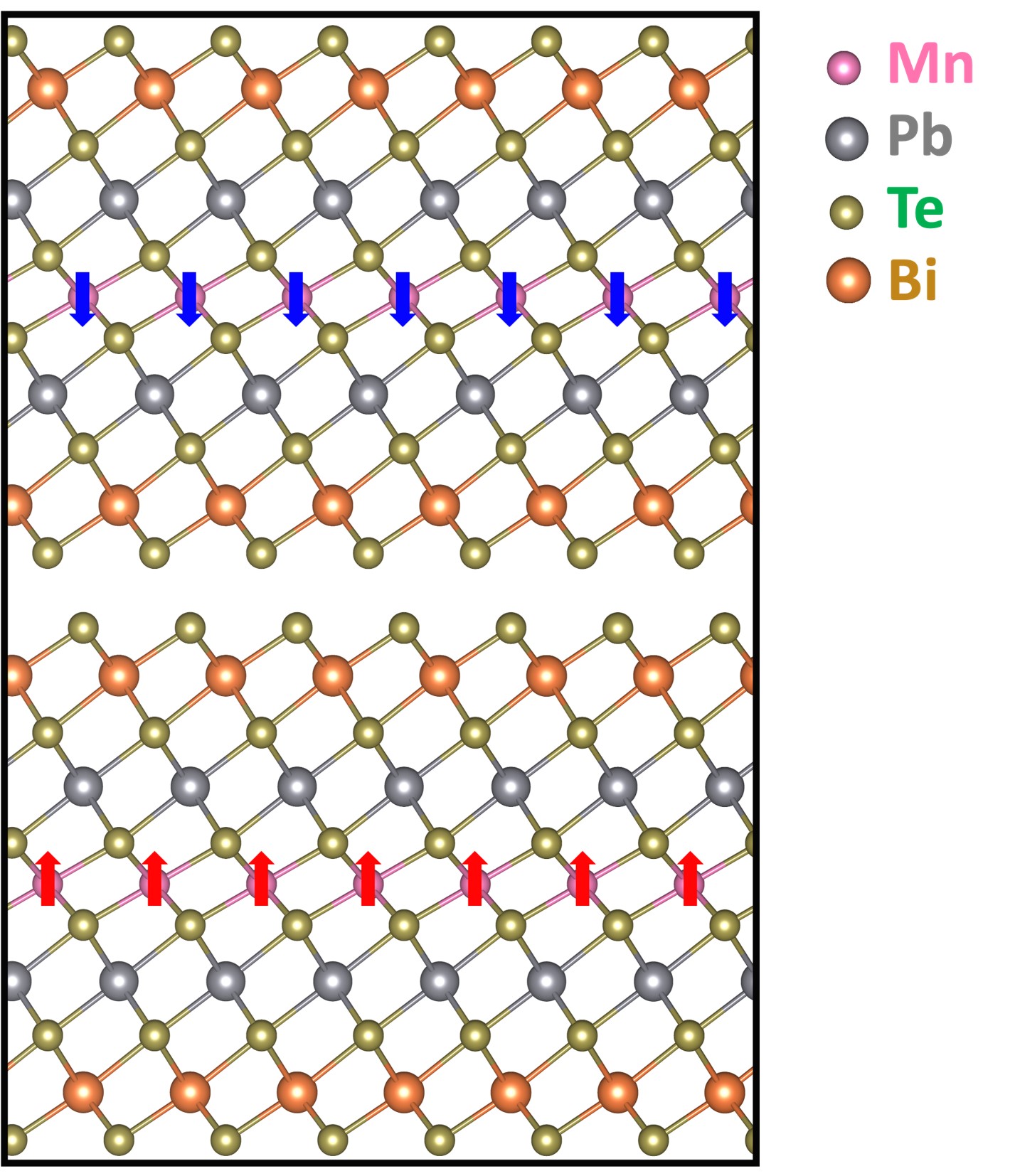}
\caption{Crystal structure of the hypothetical MnPb$_2$Bi$_2$Te$_6$ material, MPBT, where the Mn-Te layer is first flanked by two Pb-Te layers, followed by two Bi-Te layers. Pink, gray, green, and orange balls represent Mn, Pb, Te, and Bi atoms. Red and blue arrows represent up and down spin orientation along the out-of-plane direction relative to the MPBT plane.}
\label{fig:structure}
\end{figure}
The MPBT structure, proposed in recent theoretical studies~\cite{gao2022intrinsic,gao2022intrinsic2}, is formed by stacking undecuple layers consisting of Mn, Pb, Bi, and Te, with vdW interactions coupling the layers together. As shown in Figure~\ref{fig:structure}, this results in a longer $c$-axis compared to the MBT structure. There are two possible stacking sequences for the MPBT undecuple layer: Te-Bi-Te-Pb-Te-Mn-Te-Pb-Te-Bi-Te and Te-Pb-Te-Bi-Te-Mn-Te-Bi-Te-Pb-Te, which is similar to Pb${2}$Bi${2}$Te$_{5}$, a material experimentally found to have two different stacking sequences~\cite{petrov1970electron,chatterjee2015solution}. In our study, we chose the MPBT structure with the atomic sequence Te-Bi-Te-Pb-Te-Mn-Te-Pb-Te-Bi-Te, which has lower energy than the other sequence by 369 meV per formula unit.
The magnetic properties of MPBT were also investigated in our study. We found that the A-type antiferromagnetic (AFM) magnetic ordering is energetically favored over the out-of-plane ferromagnetic (FM) ordering, regardless of the lattice constants, number of $k$-points, Hubbard $U$ values, and spin-orbit coupling effects (see Table S1 in Supporting Information~\cite{SI}). The energy difference between the AFM and FM states is as small as $\sim$0.4 meV per formula unit, which is much smaller than the energy scale of interest in our study, indicating that our results are not significantly affected by the magnetic state of MPBT. Previous studies reported that the FM order is preferred over the AFM order for the same atomic sequence as in our study. Still, we confirmed that the AFM state is energetically favored for the PBE+D3-optimized structure used in our study.
Lastly, we confirmed that the MPBT structure is dynamically stable, as evidenced by the absence of imaginary frequencies in the phonon spectrum (see Figure S1 in Supporting Information), which is important for the finite-temperature analysis presented later.

\begin{table}[t]
\begin{center}
\scalebox{0.7}{
\begin{tabular}{c||c|c|c|c||c}
\hline\hline
 & ~$U=0$ eV~ & ~$U=3$ eV~ & ~$U=3.5$ eV~ & ~$U=4$ eV~ & ~$U=3.5$ eV + SOC~ \tabularnewline
\hline
  ~(i) MnBi$_{2}$Te$_{4}$ + 2PbTe $\rightarrow$ MnPb$_{2}$Bi$_{2}$Te$_{6}$~ & 7.1 & 2.9 & 2.4 & 1.9 & 1.7 \tabularnewline
  ~(ii) Pb$_{2}$Bi$_{2}$Te$_{5}$ + MnTe $\rightarrow$ MnPb$_{2}$Bi$_{2}$Te$_{6}$~ & 22.8 & -1.9 & -4.7 & -7.3 & -5.1 \tabularnewline
  ~(iii) MnTe + Bi$_{2}$Te$_{3}$ + 2PbTe $\rightarrow$ MnPb$_{2}$Bi$_{2}$Te$_{6}$~ & 21.5 & -3.3 & -6.1 & -8.7 & -7.4 \tabularnewline
  ~(iv) MnTe + Bi$_{2}$Te$_{3}$ $\rightarrow$ MnBi$_{2}$Te$_{4}$~ & 22.7 & -9.8 & -13.4 & -16.7 & -14.3 \tabularnewline
\hline\hline
\end{tabular}}
\caption{The enthalpies of formation of the possible reactions (i)-(iii) for MPBT and the standard reaction (iv) for MBT depend on different values of Hubbard $U$ and the inclusion of SOC on top of the PBE+D3 calculations. The energies are in the unit of meV/atom.}
\label{0K_energy_table}
\end{center}
\end{table}

To assess the reaction stability of MPBT, we calculated the energy differences between the reactants and products for three different reactions: (i) MnBi$_{2}$Te$_{4}$ + 2PbTe $\rightarrow$ MnPb$_{2}$Bi$_{2}$Te$_{4}$, (ii) Pb$_{2}$Bi${2}$Te${5}$ + MnTe $\rightarrow$ MnPb$_{2}$Bi$_{2}$Te$_{4}$, and (iii) MnTe + Bi$_{2}$Te$_{3}$ + 2PbTe $\rightarrow$ MnPb$_{2}$Bi$_{2}$Te$_{4}$. We also calculated the energy difference for the standard reaction of MBT, (iv) MnTe + Bi$_{2}$Te$_{3}$ $\rightarrow$ MnBi$_{2}$Te$_{4}$. The results of these calculations are presented in Table~\ref{0K_energy_table}. To account for the electronic correlation of the 3$d$ orbitals of Mn, we employed three different values of Hubbard $U$ (3, 3.5, and 4~eV). Also, we considered the impact of spin-orbit coupling (SOC) effects for the case of $U$ = 3.5~eV, which was benchmarked from diffusion Monte Carlo calculations as the best value to describe the magnetic properties of MBT~\cite{bennett2022magnetic}. We assumed antiferromagnetic ordering for the magnetic materials of MnTe, MBT, and MPBT. We determined the stability of the products based on the sign of the reaction energy (negative for stable and positive for unstable products).

We performed DFT($U$=0) calculations to evaluate the thermodynamic stability of MPBT and found that the resulting reaction energies were relatively small, with a maximum energy scale of only 23~meV/atom. This suggests that finite-temperature effects will be an essential factor in determining the overall stability of MPBT. We also examined the effects of Hubbard $U$ and SOC on the reaction energies and found that while both factors played a role, Hubbard $U$ had a greater impact. Specifically, the inclusion of Hubbard $U$ resulted in different signs of the reaction energies for reactions (ii) and (iii) compared to the standard reaction for MBT (iv). In addition, we observed that the reaction energies for reaction (i), which did not involve the MnTe compound, were the least sensitive to variations in $U$. This is likely due to the similar chemical environments around Mn in MBT and MPBT. However, when the MnTe compound was involved in the reaction, the structure and bonding were distinct from MBT and MPBT, leading to greater sensitivity to onsite correlation ($U$) and SOC. Ultimately, by including onsite correlation ($U$) and SOC, we found that MPBT was within 2~meV/atom of stability relative to MBT at $T=0$~K.

\begin{figure}[t]
\centering
\includegraphics[width=6.0in]{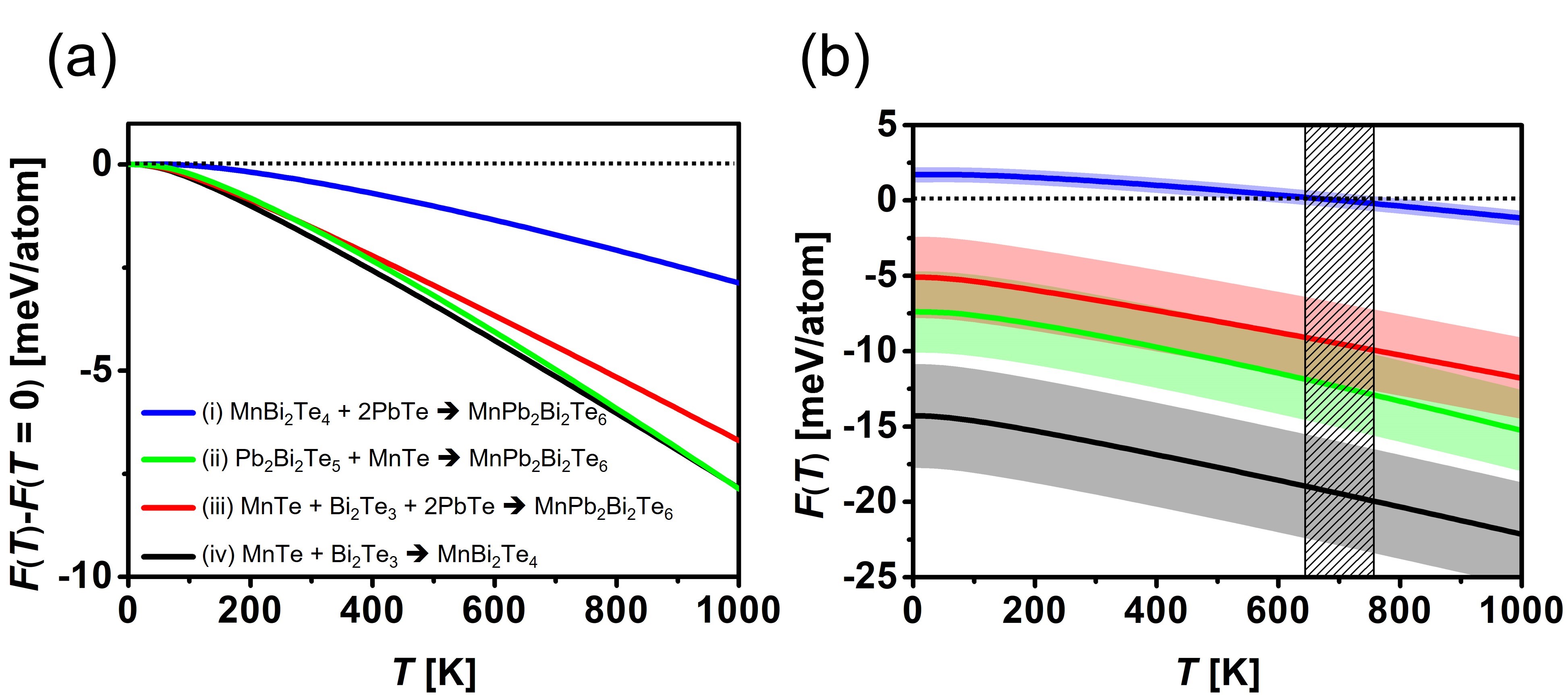}
\caption{Thermodynamic stability of the various reaction for MPBT. (a) Change Helmholtz free energy at finite temperatures for the possible reactions listed in Table~\ref{0K_energy_table}. (b) Full Helmholtz free energy of the same reactions, including contributions from the zero-temperature total energy of PBE+D3+$U$(3.5 eV)+SOC and zero point motion. The patterned-shaded region corresponds to temperatures commonly visited during the synthesis of MBT~\cite{lee2013crystal,zeugner2019chemical}. In contrast, the colored shaded ones represent computational spread using different values of Hubbard $U$, namely, the difference between the results of $U$ = 3 and 4 eV tabulated in Table~\ref{0K_energy_table}.
}
\label{fig:free_energy}
\end{figure}

The intricate energy scale associated with the 0~K energetics and their sensitivity to on-site energy $U$ values and SOC effects necessitates the calculation of Helmholtz free energy. This method enables us to discuss the thermodynamic stability of MPBT based on the phonon spectra for the materials involved in reactions (i)-(iv), depicted in Figure S1 of the Supporting information~\cite{SI}. Figure~\ref{fig:free_energy}(a) displays a decline in the reaction free energy with rising temperature for all reactions, indicating that the products become increasingly stable. However, the energy gained from the phonons' finite-temperature effects is negligible, with less than 7 and 5~meV/atom for the MBT and MPBT reactions, respectively, within the temperature range from 0 to 1000~K. Moreover, the direct reaction between MBT and MPBT (reaction (i)) exhibits the weakest stabilizing effect, causing the two structures to be closely competitive in energy at MBT formation temperatures. As a result, the phonon temperature effects do not provide sufficient advantages to stabilize MPBT over MBT. This observation aligns with the metastable formation of MBT in a narrow temperature range~\cite{lee2013crystal,zeugner2019chemical}, as depicted by the pattern-shaded area in Figure~\ref{fig:free_energy}(b), implying that MPBT formation may encounter similar difficulties. However, reactions (ii) and (iii) could prove beneficial for MPBT formation, as shown in Figure~\ref{fig:free_energy}(b). The figure presents the full Helmholtz free energies, with the zero-temperature contributions obtained from PBE+D3+$U$(3.5~eV)+SOC calculations. The colored shaded regions represent their corrections, and the reaction energy difference between the results of $U$ = 3 and 4~eV was added to the finite-temperature data to display the sensitivity observed in Table~\ref{0K_energy_table}. Our findings reveal that MPBT is energetically favored over binary materials (red curves) and a system comprising Pb${2}$Bi${2}$Te${5}$ and MnTe (green curves), indicating that MPBT would remain energetically preferred even if Pb${2}$Bi${2}$Te${5}$ were segregated into the binary PbTe or Bi${2}$Te${3}$. In contrast, the reaction involving MBT is energetically disadvantageous for forming MPBT, even at high temperatures of 1000~K.

\begin{figure}
\centering
\includegraphics[width=5.4in]{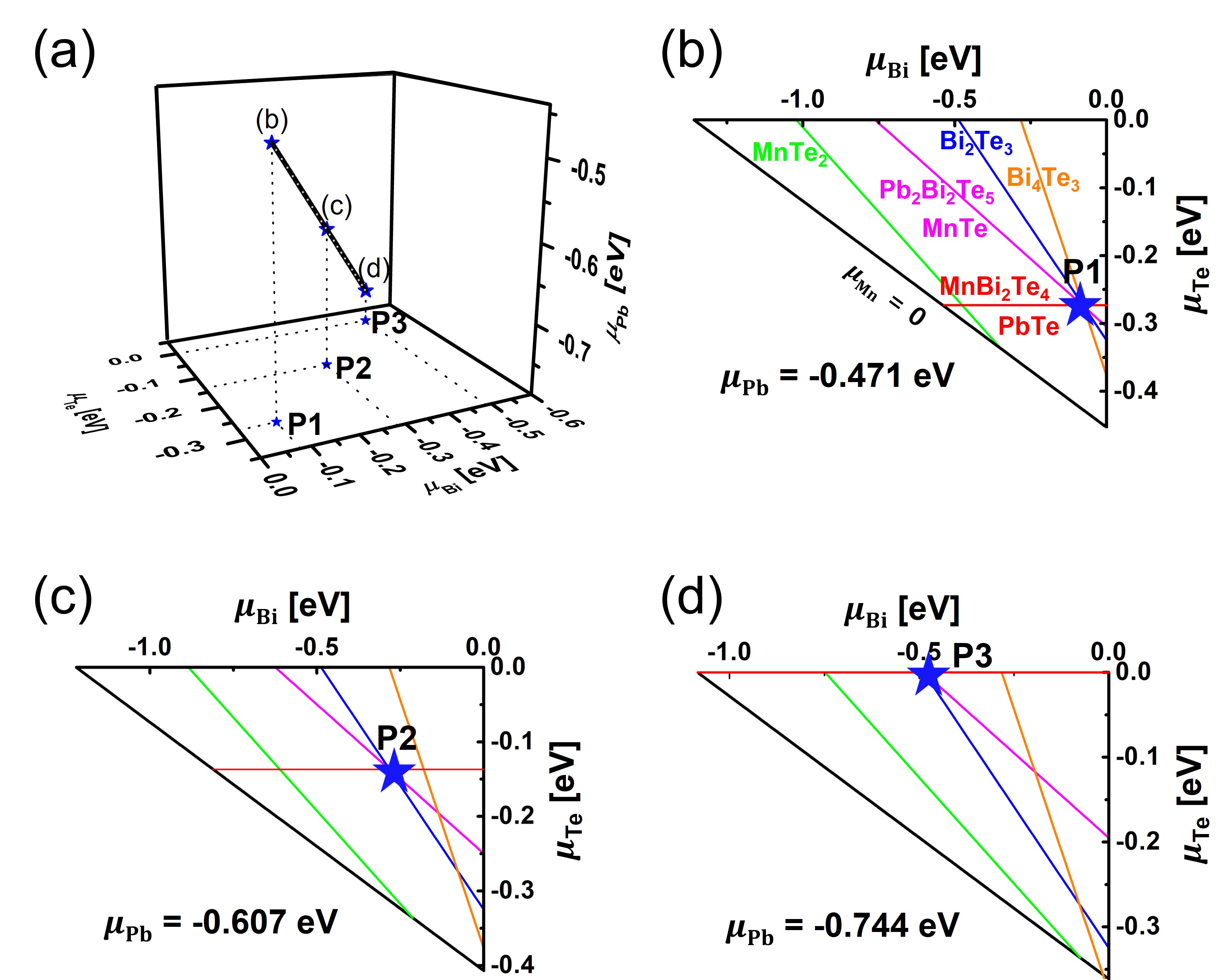}
\caption{Phase diagram of relevant binary and ternary compounds in the space constituted with $\mu_{\text{Bi}}$, $\mu_{\text{Te}}$ and $\mu_{\text{Pb}}$ computed with the PBE+D3+$U$(3.5 eV)+SOC scheme, which allows $\mu_{\text{Pb}}$ to be within the range of -0.744 $\sim$ -0.471 eV. On the ($\mu_{\text{Bi}}$, $\mu_{\text{Te}}$) plane perpendicular to the $\mu_{\text{Pb}}$ axis, the stable region appears as the intersection between red and magenta lines and evolves as $\mu_{\text{Pb}}$ changes from its maximum value of (a) -0.471 eV to the minimum of (c) -0.744 eV, passing through the intermediate value of (b) -0.607 eV. The intersections in (a) and (c) correspond to the Te-poor/cation-rich and Te-rich/cation-poor limits, respectively. The line connecting two endpoints in the three-dimensional space of $\mu_{\text{Pb}}$, $\mu_{\text{Bi}}$ and $\mu_{\text{Te}}$ result in the linear stable regions to be utilized for the evaluation of the defect formation energies.}
\label{fig:phase_diag_soc} 
\end{figure}

In light of pursuing its intrinsic properties, we now assessed how robust MPBT is to non-stoichiometry and intersite mixing by computing the formation energies of intrinsic substitutional and vacancy defects via  
\begin{equation}
 \Delta (q,\epsilon_{f}) = E_{D} - E_{h} - \sum_{i}n_{i}(\mu_{i}+\mu_{i}^{\text{bulk}}) +q(\epsilon_{\text{VBM}} + \epsilon_{f}),
 \label{eqn:areaset}
\end{equation}  
where $E_{D}$ and $E_{h}$ denote the PBE+$U$ total energies of defective and pristine structures, $n_{i}$ represents the difference in the number of defects for atomic species denoted by $i$, and the corresponding chemical potential is represented by $\mu_{i}$. The defect formation energy was calculated by first evaluating the chemical potentials for the atomic species. As we only considered the neutral case ($q = 0$), we used the same procedure as described in Ref.~\cite{du2021tuning} for the analogous system MBT.
A set of inequalities and one equality allows us to determine chemical potentials under the equilibrium growth condition as presented below,
 \begin{align}
 \mu_{\text{Mn}} + 2\mu_{\text{Pb}} + 2\mu_{\text{Bi}} + 6\mu_{\text{Te}} = \Delta H(\text{Mn}\text{Pb}_{2}\text{Bi}_{2}\text{Te}_{6}),\\
 \mu_{\text{Mn}} \le 0, ~\mu_{\text{Mn}} \le 0, ~\mu_{\text{Mn}} \le 0, ~\mu_{\text{Mn}} \le 0,\\
 \mu_{\text{Mn}} + 2\mu_{\text{Bi}} + 4\mu_{\text{Te}} \le \Delta H(\text{Mn}\text{Bi}_{2}\text{Te}_{4}),\nonumber\\
 2\mu_{\text{Pb}} + 2\mu_{\text{Bi}} + 5\mu_{\text{Te}} \le \Delta H(\text{Pb}_{2}\text{Bi}_{2}\text{Te}_{5}),\nonumber\\
 \mu_{\text{Mn}} + \mu_{\text{Te}} \le \Delta H(\text{MnTe}), ~\mu_{\text{Mn}} + 2\mu_{\text{Te}} \le \Delta H(\text{MnTe}_{2}),\nonumber\\
 \mu_{\text{Bi}} + \mu_{\text{Te}} \le \Delta H(\text{BiTe}), ~2\mu_{\text{Bi}} + 3\mu_{\text{Te}} \le \Delta H(\text{Bi}_{2}\text{Te}_{3}),\nonumber\\
 4\mu_{\text{Bi}} + 3\mu_{\text{Te}} \le \Delta H(\text{Bi}_{4}\text{Te}_{3}), ~8\mu_{\text{Bi}} + 9\mu_{\text{Te}} \le \Delta H({\text{Bi}_{8}\text{Te}_{9}}),\nonumber\\
 \mu_{\text{Pb}} + \mu_{\text{Te}} \le \Delta H(\text{PbTe})\nonumber.
 \label{eqn:defect_formation}
 \end{align} 
To ensure the stability of MPBT during growth, a thermodynamic constraint is imposed that reduces the independent variables to just $\mu_{\text{Bi}}$, $\mu_{\text{Te}}$ and $\mu_{\text{Pb}}$. A set of inequalities is also included to prevent the formation of ternary (MnBi$_{2}$Te$_{4}$ and Pb$_{2}$Bi$_{2}$Te$_{5}$) and binary phases (MnTe, MnTe$_{2}$, BiTe, Bi$_{2}$Te$_{3}$, Bi$_{4}$Te$_{3}$, Bi$_{8}$Te$_{9}$ and PbTe), as well as elemental bulk solids for each species during growth. Following the approach taken for MBT~\cite{du2021tuning}, we assume all enthalpies of formation for the reactions listed in table~\ref{0K_energy_table} to be zero due to the challenging synthesis conditions for MPBT. As a result, the phase boundary planes between ternary and binary phases (MBT/PbTe and Pb$_{2}$Bi$_{2}$Te$_{5}$/MnTe) are identical to each other. This leads to the stable range of chemical potentials appearing as a line segment in the three-dimensional space defined by the $\mu{\text{Bi}}$, $\mu_{\text{Te}}$, and $\mu_{\text{Pb}}$ axes. 

Whether the chemical potential range is stable or not ends up depending on the existence of an intersection between two lines (see red and magenta lines in Figure~\ref{fig:phase_diag_soc}) on the ($\mu_{\text{Bi}}$,$\mu_{\text{Te}}$) plane, perpendicular to the $\mu_{\text{Pb}}$ axis, within the range constructed from the above inequalities. Thus, the connection of the intersections for each plane resulting from different $\mu_{\text{Pb}}$ appears as the line segment, as described in Figure~\ref{fig:phase_diag_soc}. It represents the evolution of the stable point of the chemical potentials in the ($\mu_{\text{Bi}}$,$\mu_{\text{Te}}$) plane for decreasing $\mu_{\text{Pb}}$ from its maximum ($\mu_{\text{Pb}}=-0.471$~eV) to the minimum ($\mu_{\text{Pb}}=-0.744$~eV), computed with PBE+D3+$U$(3.5~eV)+SOC. Once the intersection of ($\mu_{\text{Bi}}$,$\mu_{\text{Te}}$) is determined for each value of $\mu_{\text{Pb}}$ using the thermodynamic constraint of equations (3) and (4), one can find the set of ($\mu_{\text{Pb}}$, $\mu_{\text{Bi}}$, $\mu_{\text{Te}}$, $\mu_{\text{Mn}}$) using equation (3). Following this procedure, we determine the stable region of chemical potentials under the growth condition for MPBT to be (-0.471, -0.078, -0.273, 0.919) eV $\sim$ (-0.744, -0.488, 0.000, -1.193) eV, which corresponds to Te-poor/cation-rich and Te-rich/cation-poor limits, respectively.

\begin{figure}[t]
\centering
\includegraphics[width=6.0in]{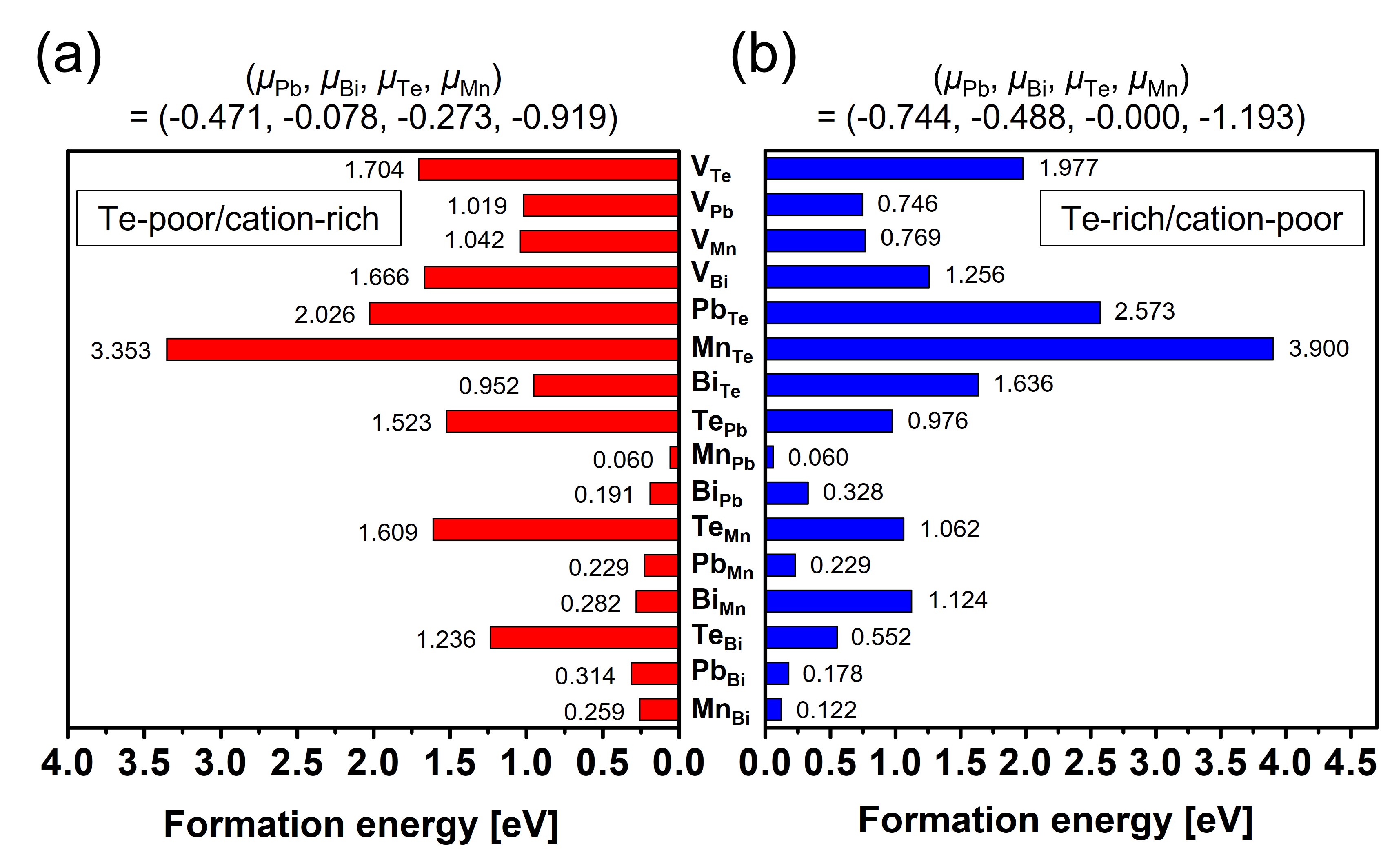}
\caption{Formation energies of neutral point defects in MPBT in the eV unit at the endpoints of the linear stability region in the phase diagram. The chemical potentials ($\mu_{\text{Pb}}$, $\mu_{\text{Bi}}$, $\mu_{\text{Te}}$, $\mu_{\text{Mn}}$) in (a) and (b) correspond to the endpoints of (a) Te-poor/cation-rich and (c) Te-rich/cation-poor limits in Figure~\ref{fig:phase_diag_soc}, respectively.
The computations were done with PBE+D3+$U$(3.5 eV)+SOC, and V$_{\text{atom}}$ denotes the vacancy defects for the given atomic site. The numerical values are presented for each defect.}
\label{fig:defect}
\end{figure}
Figure~\ref{fig:defect} displays the formation energies of various point defects that may arise during the synthesis of MPBT at the two endpoints of the stable line. We only considered point defects of neighboring antisites and vacancies. All computed defect formation energies are positive, indicating some stability against intersite mixing. The formation of vacancy and antisite defects between Te and cationic atoms is predicted to be significantly suppressed compared to other defects. Unlike the case of MBT ~\cite{du2021tuning}, where Mn$_{\text{Bi}}$ and Bi$_{\text{Mn}}$ are the most prevalent, in MPBT, their formation is not energetically most preferred at both Te-poor/cation-rich and Te-rich/cation-poor limits. Additionally, the formation energies of Pb-Bi and Pb-Mn antisite defects are computed to be small, indicating that we can expect the greatest mixing between these sub-lattices.

Among all defects studied, Mn$_{\text{Pb}}$ emerged as the most stable at both limits. This is due to the constant difference of $\mu_{\text{Pb}}-\mu_{\text{Mn}}$, which results in the defect formation energies for Mn$_{\text{Pb}}$ and Pb$_{\text{Mn}}$ being independent of the chemical potentials. It's worth noting that the formation energies for Mn$_{\text{Pb}}$ at both limits are approximately 60~meV, which is in the same energy range as the heat bath during synthesis. This suggests that MPBT may be susceptible to the formation of Mn$_{\text{Pb}}$ during synthesis at high temperatures due to an unfavorable structural arrangement of Te-Pb-Te in MPBT. PbTe crystallizes into the face-centered cubic structure, with a different relative atomic position between Pb and Te compared to that of MPBT, which belongs to the hexagonal crystal family. This leads to significant energy costs due to compressive strain on PbTe. Therefore, the system can be stabilized by replacing Pb atoms with a neighboring cationic atom that has a smaller ionic radius than Pb, such as Mn in our case. We speculate that aligning the atomic arrangement of reactants and product, along with minimizing in-plane lattice mismatch, could lead to the formation of TIs with lower concentrations of antisite defects.

Furthermore, we point out that the estimated chemical potentials and defect formation energies also showed some sensitivity to the values of Hubbard $U$ and the inclusion of SOC as presented in Figures S2 and S3 of Supporting information~\cite{SI}. With Hubbard $U$ held fixed near the physical value of 3.5 eV~\cite{du2021tuning,bennett2022magnetic,lupke2022local,gao2022intrinsic,gao2022intrinsic2}, the error in the defect formation energies due to the exclusion of SOC, which is estimated by taking the difference between PBE+D3+$U$(3.5 eV) and PBE+D3+$U$(3.5 eV)+SOC results, is 0.44 eV at maximum. On the other hand, the maximum variability in the defect formation energies when using Hubbard $U$ in the range between 3 and 4 eV is estimated to be 0.15 eV, lower than the error incurred by excluding SOC by a factor of $\sim$ 3. It means that SOC is a more crucial factor in estimating defect formation, which also calls for including SOC in similar works. When estimating the enthalpy of formation, it seems desirable to have both the SOC and the Hubbard $U$ for the observed sensitivity.
The tiny formation energy of $\text{Mn}_{\text{Pb}}$ and its sensitivity to the computational schemes lead us to conclude that MPBT is likely to host substantial intersite mixing if synthesized according to our estimation of the intersite mixing concentration as in Figure S4 of Supporting Information~\cite{SI}. The high entropy of mixing might, however, aid in stabilizing the formation of MPBT in a disordered state. Large Pb/Mn mixing may significantly reduce the strain in the Pb and Mn plains.

\section{Conclusion}
We assessed of the thermodynamic stability of MPBT for different reactions, as well as its resistance to defect formation and lattice vibrations. By analyzing the enthalpies of formation and Helmholtz free energy, we found that MPBT becomes competitive with MBT due to the energy gain from finite temperature. We also estimated the formation energy for several intrinsic point defects and found that Mn$_{\text{Pb}}$ is the most likely defect to form. The high energy cost induced by significant compressive strain at the PbTe layer suggests that the alignment of the crystal symmetry and lattice mismatch is crucial in synthesizing TIs with minimal defects. Our qualitative conclusions are insensitive to computational details, but some critical quantities, such as the formation energy of Mn$_{\text{Pb}}$ are close to the stabilization of MPBT and depend on the computational method used. We conclude that MPBT is on the brink of stability in terms of thermodynamics and defect formation, similar to MBT. Therefore, caution is necessary when proposing new TI materials based on {\it ab  initio} calculations without a thorough analysis that accounts for sources of computational sensitivity and competing physical effects. Our study highlights the importance of systematically analyzing the stability of TIs against multiple factors and developing synthesis strategies for desirable magnetic TI materials with robust stability.

\section{Acknowledgements}
\label{sec:acknowledgments}

The authors would like to acknowledge useful discussions with Jiaqiang Yan on the synthesis of new compounds.
Work by S.H.K. (calculations, writing) was supported by the U.S. DOE, Office of Science, National Quantum Information Science Research Centers, Quantum Science Center (S.H.K). Work by J.A. (calculations, writing) and J.T.K. (mentorship, writing) was supported by the U.S. Department of Energy, Office of Science, Basic Energy Sciences, Materials Sciences and Engineering Division, as part of the Computational Materials Sciences Program and Center for Predictive Simulation of Functional Materials.  Work by M.H.D., M.Y., and F.A.R. (original idea, mentorship, writing) was supported by the U.S. Department of Energy, Office of Science, Basic Energy Sciences, Materials Sciences and Engineering Division.

An award of computer time was provided by the Innovative and Novel Computational Impact
on Theory and Experiment (INCITE) program. This research used resources of the Oak Ridge Leadership
Computing Facility, which is a DOE Office of Science User Facility supported under Contract DE-
AC05-00OR22725. This research also used resources of the National Energy Research Scientific Computing Center (NERSC), a U.S. Department of Energy Office of Science User Facility operated under Contract No. DE-AC02-05CH11231.

\section*{\large References}
\bibliography{arXiv.bbl}

\end{document}